\documentclass[usenatbib,referee]{mn2e}
\usepackage{graphicx}
\usepackage{epsfig}
\usepackage{color}

\title[The violent merger scenario for the progenitors of SNe Ia]
{The violent white dwarf merger scenario for the progenitors of type Ia supernovae}
\author[D.-D. Liu et al.]
{ D.-D. Liu$^{\rm 1,2,3}$\thanks{E-mail:liudongdong@ynao.ac.cn}, B. Wang$^{\rm 1,2}$\thanks{E-mail:wangbo@ynao.ac.cn}, Ph. Podsiadlowski$^{\rm 4}$\thanks{E-mail:podsi@astro.ox.ac.uk} and Z. Han$^{\rm 1,2}$\thanks{E-mail:zhanwenhan@ynao.ac.cn} \\
$^1$Yunnan Observatories, Chinese Academy of Sciences, Kunming 650216, China\\
$^2$Key Laboratory for the Structure and Evolution of Celestial Objects, Chinese Academy of Sciences, Kunming 650216, China\\
$^3$University of Chinese Academy of Sciences, Beijing 100049, China\\
$^4$Department of Astronomy, Oxford University, Oxford OX1 3RH}

\begin{document}
\date{}
\pagerange{\pageref{firstpage}--\pageref{lastpage}} \pubyear{2016}
\maketitle

\label{firstpage}

\begin{abstract}\label{0. abstract}
Recent observations suggest that some type Ia supernovae (SNe Ia)
originate from the merging of two carbon-oxygen white dwarfs (CO
WDs). Meanwhile, recent hydrodynamical simulations have indicated that
the accretion-induced collapse may be avoided under certain conditions
when double WDs merge violently. However, the properties of SNe Ia
from this violent merger scenario are highly dependent on a particular
mass-accretion stage, the so-called WD + He subgiant channel, during
which the primary WD is able to increase its mass by accreting He-rich
material from a He subgiant before the systems evolves into a double
WD system. In this article, we aim to study this particular
evolutionary stage systematically and give the properties of violent
WD mergers. By employing the Eggleton stellar evolution code, we
followed a large number of binary calculations and obtained the
regions in parameter space for producing violent mergers based on the
WD + He subgiant channel. According to these simulations, we found
that the primary WDs can increase their mass by $\sim0.10-0.45\,\rm
M_{\odot}$ during the mass-accretion stage. We then conducted a series
of binary population synthesis calculations and found that the
Galactic SN Ia birthrate from this channel is about
$0.01-0.4\times10^{\rm -3}\rm yr^{\rm -1}$. This suggests that the
violent WD mergers from this channel may only contribute to
$\sim0.3\%-10\%$ of all SNe Ia in our Galaxy. The delay times of
violent WD mergers from this channel are $\ge 1.7\rm Gyr$,
contributing to the SNe Ia in old populations. We also found that the
WD + He subgiant channel is the dominent way for producing violent WD
mergers that may be able to eventually explode as SNe Ia.
\end{abstract}


\section{Introduction} \label{1. Introduction}
Type Ia supernovae (SNe Ia), which are defined as SN explosions
without H and He lines but with strong SiII absorption lines in their
spectra, have been successfully used as standard distance indicators
in cosmological studies of dark energy (e.g. Riess et al. 1998;
Perlmutter et al. 1999). There is a theoretical consensus that SNe Ia
result from thermonuclear explosions of carbon-oxygen white dwarfs (CO
WDs) in close binaries (Hoyle \& Fowler 1960). According to the nature
of the mass donor, two widely accepted models have been proposed,
i.e. the single-degenerate (SD) model (e.g. Whelan \&Iben 1973;
Hachisu, Kato \& Nomoto 1996; Li \& van den Heuvel 1997; Langer et al.
2000; Han \& Podsiadlowski 2004; Meng et al. 2009; Wang, Li \& Han 2010) and the
double-degenerate (DD) model (e.g. Webbink 1984; Iben \& Tutukov 1984;
Nelemans et al. 2001; Toonen, Nelemans \& Portegies 2012). The
difference between these two models is whether the mass donor is a
non-degenerate star (a main-sequence star, a red-giant star or a
helium star; the SD model) or another WD merging with the primary WD
(the DD model). However, there is still no conclusive evidence to
support any progenitor models of SNe Ia. Recent studies suggested that
more than one progenitor models may be required to reproduce the
observational diversities of SNe Ia (see Podsiadlowski et al. 2008;
Howell et al. 2011; Wang \& Han 2012; Maoz, Mannucci \& Nelemans
2014).

Some recent observations seem to favor the DD model: for example, the
lack of H and He lines in the nebular spectra of most SNe Ia
(e.g. Leonard 2007; Ganeshalingam, Li \& Filippenko 2011), no
conclusive proof for the existence of surviving companions
(e.g. Badenes et al. 2007; Graham et al. 2015), the lack of radio
emission (e.g. Hancock, Gaensler \& Murphy 2011; Horesh et al. 2012)
and the observed evidence for the existence of some super-luminous
events (e.g. Howell et al. 2006; Hichen et al. 2007; Scalzo et
al. 2010).\footnote{Some of these observed clues may also be explained
  by the SD model after considering the spin-up/spin-down processes of
  WDs (e.g. Justham 2011; Di Stefano, Voss \& Claeys 2011; Hachisu,
  Kato \& Nomoto 2012; Wang et al. 2014).}
Meanwhile, the delay times of SNe Ia, defined as the time interval
from the star formation to the thermonuclear explosion, provide an
important observational constraint for SN Ia progenitor models. Recent
observations have suggested that the delay time distributions (DTDs)
of SNe Ia follow a power-law distribution $\sim t^{-1}$ (Totani et
al. 2008; Maoz et al. 2011; Graur et al. 2011; Barbary et al. 2012;
Sand et al. 2012), which is consistent with the results predicted by
the DD model (e.g. Ruiter, Belczynski \& Fryer 2009; Mennekens et
al. 2010). In addition, Dan et al. (2015) compared the nucleosynthetic
yields of thermonuclear explosions from WD mergers with the
observations and found that some of their models are good condidates
for type Ia events. Moreover, many double WDs have been proposed as
possible progenitor candidates of SNe Ia, e.g. Henize 2$-$428, KPD 1930 + 2752, WD
2020-425, V458 Vulpeculae, SBS 1150 + 599A and GD687, etc (Maxted,
Marsh \& North 2000; Geier et al. 2007, 2010; Napiwotzki et al. 2007;
Rodr\'iguez-Gil et al. 2010; Tovmassian et al. 2010; Santander-Garc\'ia et al. 2015). Especially, Henize 2-428 is a planetary nebula with double degenerate core that has a total mass of $\sim 1.76\,\rm M_{\odot}$, mass ratio $q\sim1$ and orbital period $\sim4.2\,\rm h$, which suggests that Henize 2-428 is a good candidate for super-chandrasekhar mass SNe Ia (e.g. Santander-Garc\'ia et al. 2015).

However, previous studies reveal that the outcome of double WD mergers
may be a neutron star resulting from an accretion-induced collapse,
rather than a thermonuclear explosion (e.g. Nomoto \& Iben 1985; Saio
\& Nomoto 1985; Timmes, Woosley \& Taam 1994). These arguments are
based on the assumption that the merging remnant consists of a hot
envelope or a thick disc, or even both upon the primary WD
(e.g. Kashyap et al. 2015). In that case, the accretion rate from the
envelope or disc may be relatively high, leading to the formation of
an oxygen-neon (ONe) WD that would collapse into a neutron star when
it approaches the Chandrasekhar limit (Nomoto \& Iben 1985; Saio \&
Nomoto 1998). We note that Yoon, Podsiadlowski \& Rosswog (2007)
argued that the accretion-induced collapse can be avoided for a certain
range of parameters and found that they would explode as a SN Ia some
$10^5\,\rm yr$ after the initial dynamical merger when considering the
rotation of the WDs.

Recently, Pakmor et al. (2010) proposed a new explosion scenario for
SNe Ia that are produced by the merging of double WDs referred to as
the violent merger scenario. In this scenario, a prompt detonation is
triggered while the merger is still ongoing, giving rise to a SN Ia
explosion (see also Pakmor 2011, 2012). Pakmor et al. (2010) found that
the violent merger of double WDs with almost equal masses of $0.9\,\rm M_{\odot}$
can provide an explanation for the formation of sub-luminous 1991bg-like
events. Pakmor et al. (2011) suggested that the minimum critical mass
ratio for double WD mergers to produce SNe Ia is about 0.8. R\"{o}pke
et al. (2012) recently argued that the violent merger model can reproduce
the observational properties of SN 2011fe. Additionally, this scenario may
also explain the formation of some super-Chandrasekhar mass SNe Ia (e.g. Moll
et al. 2014; Cody et al. 2014). For a series of recent theoretical and
observational studies of the violent merger scenario see
Taubenberger et al. (2013), Kromer et al. (2013), Fesen, H\"oflich \&
Hamilton (2015), Seitenzahl et al. (2015), Tanikawa et al. (2015),
Chakraborti, Childs \& Soderberg (2015) and Bulla et al. (2016).

Ruiter et al. (2013) recently investigated the distribution of the SN
Ia brightness based on the violent merger scenario and argued that the
theoretical peak-magnitude distribution from their calculations can
roughly reproduce their observed properties. The distribution they
obtained depends critically on a particular binary evolutionary stage
called the WD + He subgiant channel, during which the primary WD is
able to grow in mass by accreting He-rich material from their He
companion before eventually evolving into a double WD system. However,
this mass-accretion process is still poorly studied, which may
influence the binary population synthesis (BPS) results of SNe Ia
based on the violent merger scenario (see Ruiter et al. 2013).


In this paper, we systematically study the WD + He subgiant channel
for producing SNe Ia via the violent WD merger scenario and obtain
the parameter space for the progenitors of SNe Ia. We then present a
series of BPS simulations using this parameter space. In Sect.\,2, we
describe our methods for the binary evolution calculations and give
the main results. The methods and results of our BPS
simulations are presented in Sect.\,3. A detailed discussion is
provided in Sect.\,4 and finally a summary is given in Sect.\,5.

\section{binary evolution calculations}\label{binary evolution calculations}
\subsection{Criteria for violent mergers of double WDs}
We conducted a number of binary evolution simulations of WD + He star
systems in which the He star fills its Roche lobe at the He MS or
subgiant stage and transfers He-rich material onto the WD. The accreted
He-rich material is burned into C and O on the surface of the primary
WD, leading to an increase in mass of the WD. When the He shell in the
He subgiant donor is exhausted, a double WD system is
produced. Subsequently, the double WD system loses its orbital angular
momentum and eventually merges due to gravitational wave
radiation. We assume that a SN Ia explosion occurs via
the violent merger scenario if a double WD system satisfies the
following criteria:

(1) The critical mass ratio of the double WDs, $q_{\rm cr}=M_{\rm
  WD2}/M_{\rm WD1}$, is larger than 0.8, where $M_{\rm WD1}$ and
$M_{\rm WD2}$ are the mass of the massive WD and the less massive WD,
respectively (see Pakmor et al. 2011).

(2) The mass of the massive WD in the double WD system is larger than
$0.8\,\rm M_{\odot}$ (see Pakmor et al. 2010; Sim et al. 2010; Ruiter
et al. 2013).

(3) The delay times ($t$) of SNe Ia are shorter than the Hubble time. In this paper, the delay time of SNe Ia is defined as
\begin{equation}
t = t_{\rm evol} + t_{\rm acc} + t_{\rm GW},
\end{equation}
where $t_{\rm evol}$, $t_{\rm acc}$ and $t_{\rm GW}$ are the
evolutionary timescale from the primordial binaries to the formation
of the WD + He star systems, the evolutionary timescale from the
formation of the WD + He star systems to the formation of double WDs,
and the merging timescale of double WDs, respectively.  Here, $t_{\rm
  GW}$ is defined as the timescale for the double WDs to be brought
together and eventually merge through gravitational wave radiation
(e.g. Landau \& Lifshitz 1971):
\begin{equation}
{t_{\rm GW}} = 8 \times {10^7} \times \frac{{{{({M_{\rm WD1}} + {M_{\rm WD2}})}^{1/3}}}}{{{M_{\rm WD1}}{M_{\rm WD2}}}}{P^{8/3}},
\end{equation}
where $P$ is the orbital period of the WD binary in hours, $t_{\rm GW}$ is in unit of years, $M_{\rm WD1}$ and $M_{\rm WD2}$ are in unit of $\rm M_{\odot}$.

\subsection{Binary evolution code}
We employed the Eggleton stellar evolution code (Eggleton 1973; Han,
Podsiadlowski \& Eggleton 1994; Pols et al. 1995, 1998; Eggleton \&
Kiseleva-Eggleton 2002) to simulate the evolution of WD + He star
systems up to beginning of the merger phase. The initial setup and
physical assumptions in this code are similar to those in Wang et
al. (2009). We assumed that the He star models have a helium
mass fraction $Y=0.98$ and metallicity $Z=0.02$. The process of
Roche-lobe overflow (RLOF) is calculated using the boundary
condition
\begin{equation}
\dot{M}_{\rm 2}=C\max\left[0,{\left(\frac{r_{\rm star}}{r_{\rm lobe}}-1\right)^{\rm 3}}\right],
\end{equation}
where $\dot{M}_{\rm 2}$ is the mass-transfer rate, $r_{\rm star}$ is
the radius of the star that fills its Roche lobe, $r_{\rm lobe}$
is the radius of its Roche lobe and $C$ is a constant (see Han, Tout \&
Eggleton 2000). Here, we set $ C=1000\,\rm M_{\odot}\,yr^{\rm -1}$,
assuming that the lobe-filling star overflows the Roche lobe stably and as
necessary, but never too much, i.e. $(r_{\rm star}/r_{\rm lobe}-1)\le
0.001$.

According to WD models computed with constant mass-accretion rates,
Nomoto (1982) provided a critical mass-transfer rate written as
\begin{equation}
\dot{M}_{\rm cr}=7.2\times10^{\rm -6}(M_{\rm WD}/\rm M_{\odot}-0.6)\rm M_{\odot}\rm yr^{\rm -1}.
\end{equation}
We utilized the optically thick wind model when the mass-transfer rate
$\dot{M}_{\rm 2}$ is larger than $\dot{M}_{\rm cr}$ (Kato \&
Hachisu 1994; Hachisu et al. 1996). In this case, we assume that the
accreted He-rich material is burned into C and O stably at the rate of
$\dot{M}_{\rm cr}$, and that the unprocessed material is blown away from
the binary in the form of an optically thick wind. When $\dot{M}_{\rm
  st}\,<\dot{M}_{\rm 2}\,<\dot{M}_{\rm cr}$, we assume that the He
shell burns stably at the rate of $\dot{M}_{\rm 2}$ and that no stellar
wind is triggered, where $\dot{M}_{\rm st}$ is the minimum
accretion rate for stable He shell burning (see Kato \& hachisu
2004). When $\dot{M}_{\rm 2}\,<\dot{M}_{\rm st}$, the prescription of
Kato \& Hachisu (2004) is adopted for the mass accumulation process during
He shell flashes on the surface of the WD. During this He shell flash
process, the mass increase rate of the WD, $\dot{M}_{\rm WD}$, is
calculated as
\begin{equation}
\dot{M}_{\rm WD}=\eta_{\rm He}\dot{M}_{\rm 2},
\end{equation}
where $\eta_{\rm He}$ is the mass accumulation efficiency for He shell burning.

We added the prescriptions above in the Eggleton stellar evolution
code and followed the evolution of WD + He star systems. We assume
that the mass lost from these systems takes away the specific orbital
angular momentum of the accreting WD. In this work, we simulated the
evolution of about 800 WD + He star systems. The range of initial
masses of the WDs ($M_{\rm WD}^{\rm i}$) is $0.65-1.07\,\rm
M_{\odot}$; the upper mass limit, $1.07\,\rm M_{\odot}$, is the
maximum mass for a single CO WDs based on standard stellar models (see
Iben \& Tutukov 1985; Umeda et al. 1999). The range of the initial
masses of the He stars ($M_{\rm 2}^{\rm i}$) is $0.8-2.6\,\rm
M_{\odot}$, and the range of the initial orbital periods of the
binaries ($P^{\rm i}$) is $0.04-0.50\,\rm day$. We produced a
large and dense model grid, which can be employed in Monte Carlo BPS
simulations.

\subsection{The evolution of typical WD + He star systems}
In Figs\,1-2, we present a representative and an extreme example for the
binary evolution producing SNe Ia based on the violent merger
scenario. In each of these two figures, panel\,(a) shows the
$\dot{M}_{\rm He}$, $\dot{M}_{\rm WD}$ and $M_{\rm WD}$ as a function of
time, and panel\,(b) presents the evolutionary tracks of the He
stars in the Hertzsprung-Russell diagram and the evolution of the
orbital periods.

\begin{figure*}
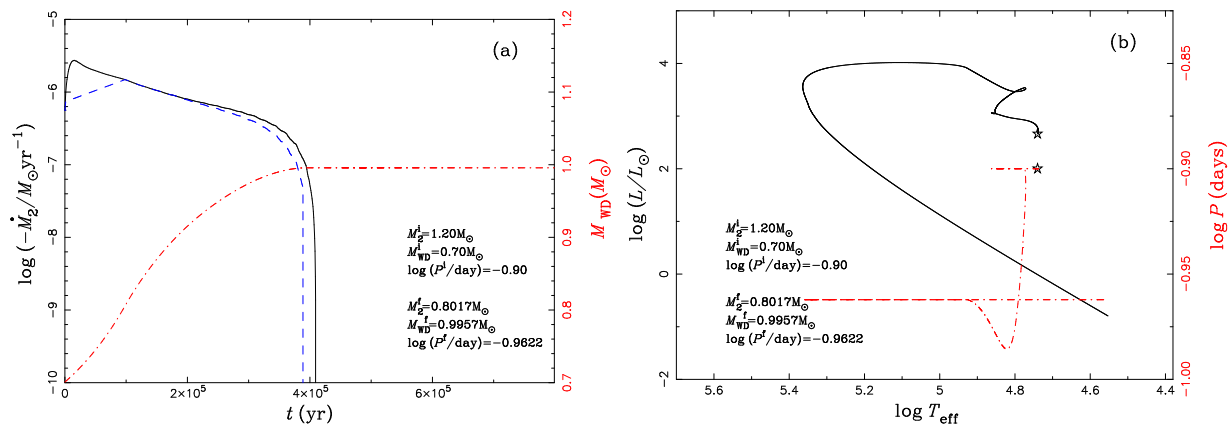

\centerline{\epsfig{file=fig1a.ps,angle=270,width=8cm}\ \ \epsfig{file=fig1b.ps,angle=270,width=8cm}} \caption{A
  representative example of binary evolution calculations that can
  produce a SN Ia through the violent merger scenario, in which the He
  star first fills its Roche lobe at the subgiant stage.  Panel (a):
  the evolution of the mass-transfer rate (solid curve), the WD
  mass-growth rate (dashed curve) and the WD mass (dash-dotted curve)
  as a function of time for the binary calculation.  Panel (b): the
  luminosity of the mass donor (solid curve) and the binary orbital
  period (dash-dotted curve) as a function of effective
  temperature. Asterisks in the right panel indicate the position
  where the simulation starts.  The initial binary parameters of WD +
  He star system and the parameters of the double WD systems at its
  formation time are also given in these two panels.}
\end{figure*}

\begin{figure*}
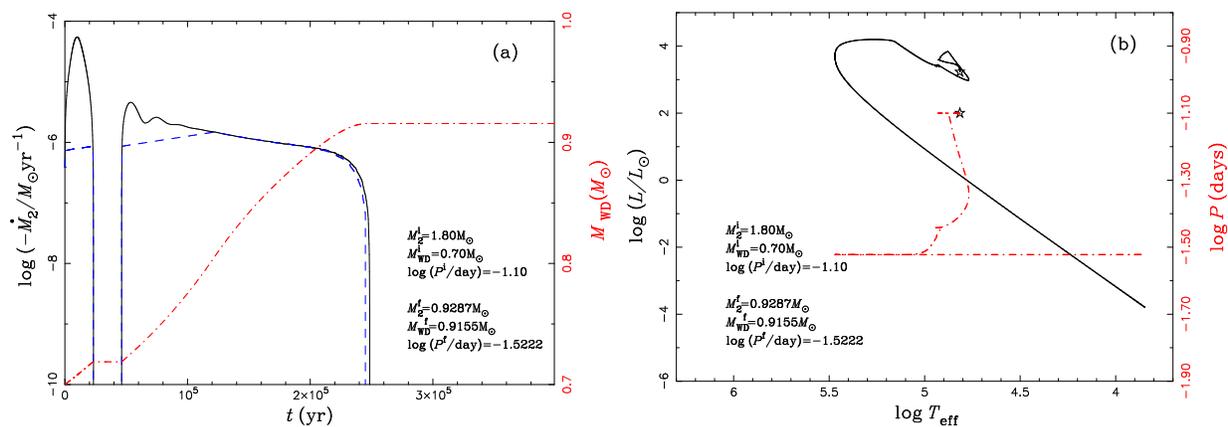

\centerline{\epsfig{file=fig2a.ps,angle=270,width=8cm}\ \ \epsfig{file=fig2b.ps,angle=270,width=8cm}} \caption{Similar
  to Fig.\,1, but for an extreme case in which the He star first fills
  its Roche lobe in the He main-sequence stage and again at the
  subgiant stage.}
\end{figure*}

Fig.\,1 shows a representative example for the evolution of a WD + He
star system in which the He star first fills its Roche lobe when it has
evolved to the subgiant stage. In this case the initial binary parameters
are ($M_{\rm WD}^{\rm i}$, $M_{\rm 2}^{\rm i}$, $\log\,P^{\rm i})=(0.7,
1.2, -0.9)$, where $M_{\rm WD}^{\rm i}$ and $M_{\rm He}^{\rm i}$ are
in units of $\rm M_{\odot}$, and $P^{\rm i}$ is in units of days. When
the He star evolves to the subgiant stage, it expands quickly and soon
fills its Roche lobe, leading to a mass-transfer phase. At the
beginning, the mass-transfer rate exceeds the critical rate
$\dot{M}_{\rm cr}$ and enters a stellar wind stage. At this stage,
part of the transferred He-rich material burns into C and O and
accumulates on the surface of the WD, while the unprocessed
part is blown away from the binary in the form of an optically thick
wind. After about $1 \times 10^5\,\rm yr$, the mass-transfer rate
drops below $\dot{M}_{\rm cr}$ and the optically thick wind stops. The
mass-transfer rate continues to decrease and drops below $\dot{M}_{\rm
  st}$ after about $1 \times 10^5\,\rm yr$. In this case, the system
is in a weak He shell flash stage and the WD still grows in
mass. After about $2 \times 10^5\,\rm yr$, the envelope of the He star
is exhausted and eventually a WD is produced. When the
double WDs are formed, the final mass of the primary WD is $M_{\rm
  WD}^{\rm f}=0.9957\,\rm M_{\odot}$, the mass of the WD produced from
the He star is $M_{\rm 2}^{\rm f}=0.8017\,\rm M_{\odot}$, and the
period of the WD binary is $\log(P^{\rm f}/\rm day)=-0.9622$. The
double WDs will merge in about $1.6\,\rm Gyr$ after their formation,
which is driven by gravitational wave radiation.

Fig.\,2 presents an extreme case in which the He star first fills its
Roche lobe at the He MS stage and again at the He subgiant stage. In
this case, the initial binary parameters are ($M_{\rm WD}^{\rm i}$, $M_{\rm
  2}^{\rm i}$, $\log\,P^{\rm i})=(0.7, 1.8, -1.1)$. Since this binary
has a short orbital period, the He star first fills its Roche lobe at
its MS stage, resulting in a stable mass-transfer process. At this
stage, $\dot{M}_{\rm 2}$ is larger than $\dot{M}_{\rm cr}$, triggering
the optically thick wind. After about $2.2 \times 10^4 \,\rm yr$, the
He star shrinks below its Roche lobe, and  mass transfer
stops. After about $2.0 \times 10^4 \,\rm yr$, the central He in the
He star is exhausted, and a CO core is formed. The He star then expands
and fills its Roche lobe again. The subsequent evolution of this
binary is similar to the case presented in Fig.\,1. When the two
WDs are form: $M_{\rm WD}^{\rm f}=0.9155\,\rm M_{\odot}$, $M_{\rm
  2}^{\rm f}=0.9287\,\rm M_{\odot}$ and $\log(P^{\rm f}/\rm
day)=-1.5222$. The double WDs will merge in about $48\,\rm Myr$ after
their formation due to gravitational wave radiation.

\subsection{Initial parameters for violent mergers of double WDs}

\begin{figure*}
\begin{center}
\epsfig{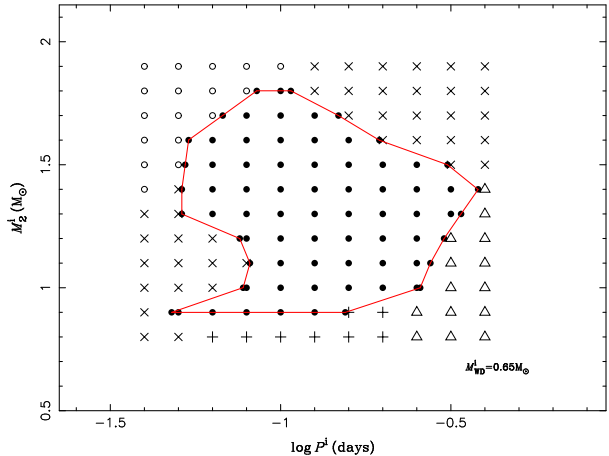}
\caption{Final results of the binary evolution calculations in the
  initial orbital period--secondary mass ($\log P^{\rm i}$, $M^{\rm
    i}_2$) plane of WD + He star systems for an initial WD mass of
  $0.65\,\rm M_{\odot}$. The filled circles denote the binaries which
  can produce violent WD mergers and eventually produce SN Ia
  explosions. The plus signs, crosses and triangles indicate that the
  formed double WD systems do not satisfy all the criteria for violent
  mergers presented in Sect.\,2.1. The open circles represent binaries
  which would enter a CE stage during the mass-transfer phase.}
\end{center}
\end{figure*}

\begin{figure*}
\begin{center}
\epsfig{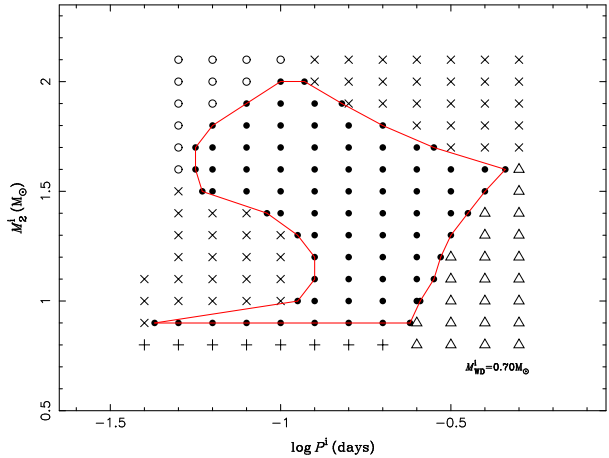}
\caption{Similar to Fig.\,3, but for an initial WD mass
of 0.7$\,\rm M_{\odot}$.}
\end{center}
\end{figure*}

\begin{figure*}
\begin{center}
\epsfig{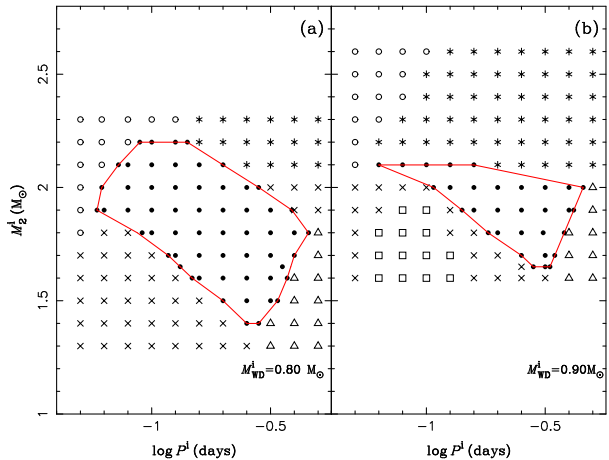}
\caption{Similar to Fig.\,3, but for initial WD masses
of 0.8 and 0.9$\,\rm M_{\odot}$. The snowflakes represent the binaries in which the He subgiants would evolve to ONe WDs but not CO WDs, and the squares denote the binaries that can produce SNe Ia through the Chandrasekar mass SD model (Wang et al. 2009).}
\end{center}
\end{figure*}

By calculating the evolution of a large number of WD + He star
systems, we obtained a large and dense model grid. In Figs\,3-6, we
present the final outcomes of the binary evolution calculations in the
initial orbital period$-$secondary mass ($\log P^{\rm i}-M^{\rm i}_2$)
plane. In these figures, the initial masses of the primordial WDs
range from $0.65$ to $1.07\,\rm M_{\odot}$. The filled circles
represent WD + He star systems which can explode as SNe Ia in their
future evolution based on the violent merger scenario. We also present
the contours of initial parameters for producing SNe Ia in these
figures and found that the primary WDs may increase their mass by
about $0.10-0.45\,\rm M_{\odot}$ by accretion from the He donors.
According to these figures, the initial masses of He
stars ($M^{\rm i}_{\rm 2}$) range from $0.9$ to $2.2\,\rm M_{\odot}$
and the initial periods of the binaries ($P^{\rm i}$) range from 1.2
to $12\,\rm h$ for producing SNe Ia. For a larger metallicity,
we expect that the grids will move to larger initial He-star masses
and larger initial orbital periods, which leads to earlier SN Ia
explosions (e.g. Wang \& Han 2010).


The WD + He star systems in Figs\,3-6 denoted by crosses, pluses and
triangles will evolve to double WDs that do not satisfy all the
criteria for violent mergers presented in Sect.\,2.1; i.e. these
systems will fail to produce SNe Ia via the violent merger
scenario. Among these binaries represented by crosses (in the
upper right regions) fail to form SNe Ia as the WDs resulting from
the He donors are too massive, leading to the mass ratios ($q$) less
than 0.8; those in the lower left regions fail to form SNe Ia as
the primary WDs are too massive, resulting in $q<0.8$. For the
binaries denoted by open circles in Figs\,3-5, mass transfer is
dynamically unstable when the He subgiants fill their Roche lobe,
leading to a common envelope (CE) stage. These binaries may also
evolve to double WDs and produce SNe Ia through the violent merger
scenario, which will be discussed further in Sect.\,3. Massive He
stars in binaries marked by snowflakes will evolve to ONe WDs but not
CO WDs; the outcome of the merger of CO WD + ONe WD systems may be a
hybrid SN, in which the ONe core may collapse and CO upon the surface
may detonate (see Dan et al. 2014). The primary WDs in binaries
indicated by squares will increase their mass to the Chandrasekhar
limit and explode as SNe Ia before the He stars evolve to the WD stage
(see Wang et al. 2009).

\begin{figure*}
\begin{center}
\epsfig{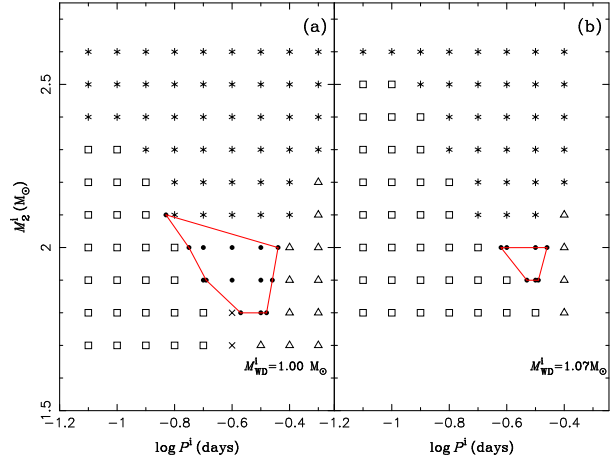}
\caption{Similar to Fig.\,3, but for initial WD masses
of 1.0 and 1.07$\,\rm M_{\odot}$.}
\end{center}
\end{figure*}

In Fig.\,7, we show the regions in the parameter space of WD + He star
systems which can lead to SNe Ia via violent mergers with various
initial WD masses (i.e. $M_{\rm WD}^{\rm i}=0.65$, 0.7, 0.8, 0.9, 1.0
and $1.07\,\rm M_{\odot}$). In this figure, it is obvious that the
contours for producing violent WD mergers move to the upper right
region as the initial WD mass increases, which follows from the
requirement of a large mass ratio ($q>0.8$). Massive He stars tend to
form massive WDs, and large orbital periods result in little material
being transferred from the He stars to the WDs. Thus, for the binaries
with massive He donors and large periods in the upper right region,
the WDs evolved from He stars could be massive enough to merge
violently with the primary WDs as the initial mass of the WDs
increases.

\begin{figure}
\begin{center}
\epsfig{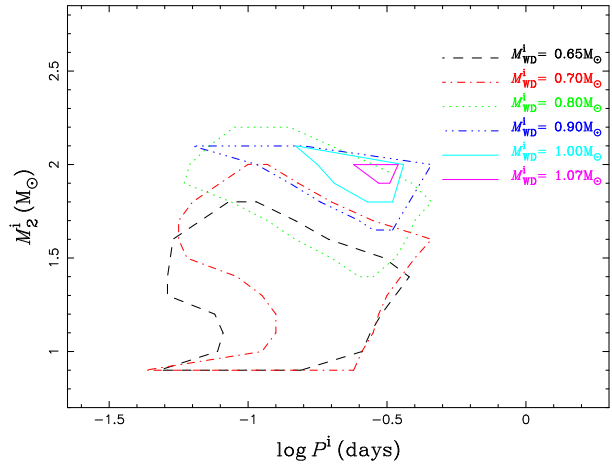}
 \caption{Regions in the orbital period-secondary mass ($\log P^{\rm i}-M_{\rm 2}^{\rm i}$) plane for WD + He star systems which can produce SNe Ia for various initial WD masses.}
  \end{center}
\end{figure}

\section{Binary population synthesis}\label{Binary population synthesis}
\subsection{BPS Methods}
Binary population synthesis, which is to follow the evolution of
millions of primordial binaries, is an important method to obtain the
birthrates and delay times of SNe Ia (e.g. Han 1998; Yungelson \&
Livio 1998; Nelemans et al. 2001).  We performed a series of BPS Monte
Carlo simulations to investigate the properties of the violent
mergers. We assume that a SN Ia explosion would occur if the initial
parameters of a WD + He star system are located in the contours of the
$\log P^{\rm i}-M_{\rm 2}^{\rm i}$ plane for the specific $M^{\rm i}_{\rm
  WD}$. The properties of the double WDs at the moment of their
formation are investigated by interpolations in the three-dimensional
grid ($M_{\rm WD}^{\rm i},M_{\rm 2}^{\rm i},\log P^{\rm i}$) of the WD
+ He star systems calculated in Sect.\,2.  Employing the rapid binary
evolution code developed by Hurley, Tout \& Pols (2002), we
simulated the evolution of binaries from their formation to the
formation of WD + He star systems with a metallicity $Z=0.02$. In each
simulation, $4\times10^7$ primordial binary samples are included.

The basic assumptions and initial parameters in our Monte Carlo
simulations are described in the following: (1) We employed the
initial mass function of Miller \& Scalo (1979) for the distribution
of the primordial primary mass ($M_{\rm 1}$). (2) For the primordial
secondary mass ($M_{\rm 2}$), we take a constant mass ratio ($q^{\rm
  '}=M_{\rm 1}/M_{\rm 2}$) distribution, i.e. $n(q^{\rm
  '})=1$. (3) All stars are considered to be members of binaries
  with circular orbits. (4) The distribution of initial orbital
  separations is assumed to be constant in $\log\,a$ for wide binaries
  and falls off smoothly for close binaries, where $a$ is orbital
  separation (Han, Tout \& Eggleton 1995):
\begin{equation}
a\cdot n(a)=\left\{
\begin{array}{lcl}
\alpha_{\rm set}, {a_{\rm 0}<a<a_{\rm 1}},\\
\alpha_{\rm set}(a/a_{\rm 0})^{\rm m}, {a\le a_{\rm 0}},
\end{array}
\right.
\end{equation}
where $\alpha_{\rm set}\approx0.07$, $m\approx1.2$, $a_{\rm 0}=10\,\rm R_{\odot}$ and $a_{\rm 1}=5.75\times10^{\rm 6}\,\rm R_{\odot}$. This distribution of orbital separations implies that about 50 percent of stellar systems have orbital periods less than 100 yr, and that there is an equal number of wide binary systems per logarithmic interval.

We simply employed a constant star formation rate (SFR) to provide a rough description of spiral galaxies, or alternatively, a delta-function SFR (i.e. a single starburst) to approximate elliptical galaxies (or star clusters). For the case of a constant SFR, the SFR is calibrated by assuming that a binary with its primary mass larger than $0.8\,\rm M_{\odot}$ is formed per year (see Iben \& Tutukov 1984; Han et al. 1995; Hurley et al. 2002). From this calibration, a SFR of $5\,\rm M_{\odot}\,\rm yr^{\rm -1}$ over the past $15\,\rm Gyr$ is obtained (see also Willems \& Kolb 2004), which is similar to the situation of our Galaxy (Yungelson \& Livio 1998; Han \& Podsiadlowski 2004). For the case of a delta-function SFR, a burst producing $10^{10}\,\rm M_{\odot}$ in stars is adopted.

The WD + He star systems originate from CE evolution (see Sect.\,3.2). However, the process of CE ejection is still unclear (e.g. Ivanova et al. 2013). The CE interaction is usually parameterized based on the relationship between the orbital energy and the binding energy. We adopted the standard energy prescription to calculate the output of the CE stage (see Webbink 1984). There exist two uncertain parameters for this prescription, i.e. the CE ejection efficiency ($\alpha_{\rm CE}$) and a stellar structure parameter ($\lambda$). The value of $\lambda$ depends on the structure and evolutionary phase of the mass donor. For simplicity, we combined $\alpha_{\rm CE}$ and $\lambda$ as a single free parameter (i.e. $\alpha_{\rm CE}\lambda$) to describe the process of the CE ejection, and set $\alpha_{\rm CE}\lambda=0.5$, 1.0 and 1.5 to examine its influence on the final results.

\subsection{Binary evolutionary channels}
There are two evolutionary channels that can produce WD + He star
systems and then form SNe Ia depending on the evolutionary phase of the
primordial primary when the primordial secondary has evolved to the He-star
phase (see Fig.\,8):

\begin{figure}
\begin{center}
\epsfig{file=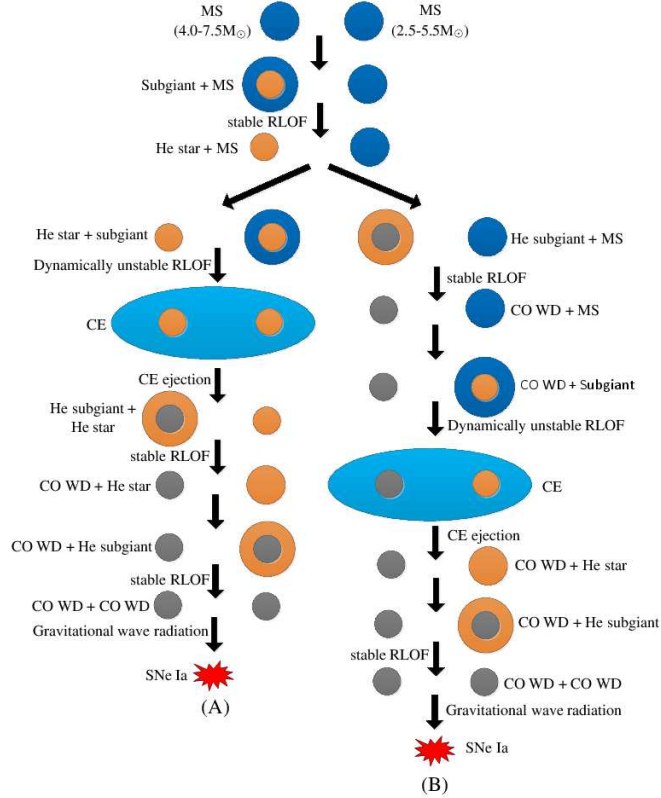,angle=0,width=9.2cm}
 \caption{Binary evolutionary channels for producing SNe Ia from the violent merger scenario.}
  \end{center}
\end{figure}

{\em Channel} A: The primordial primary first fills its Roche lobe
when it has evolved to the subgiant stage, resulting in a stable
mass-transfer process. At the end of RLOF, the primary
becomes a He star. Subsequently, the secondary evolves to the
subgiant stage and fills its Roche lobe. This mass-transfer phase is
dynamically unstable, leading to a CE process. If the CE can be
ejected, the secondary turns into a He star, and the primary evolves to
the He subgiant stage. The primary soon fills its Roche lobe again and
transfers He-rich material stably onto the surface of the
secondary. After mass transfer, the He shell of the primary will
be exhausted, and a WD + He star system will be produced. The following
evolution of the binary is similar to that presented in Figs\,1-2 and
a SN Ia would eventually be produced via the violent merger
scenario. For this channel, the primordial binaries are in the range
of $M_{\rm 1,i}\sim4.0-5.5\,\rm M_{\odot}$, $M_{\rm
  2,i}\sim2.5-4.5\,\rm M_{\odot}$ and $P^{\rm i}\sim5-330\,\rm days$.

{\em Channel} B: The evolution of this channel is similar to that of
{\em Channel} A before the primary evolves to be a He star. The
primary then continues to evolve and will fill its Roche lobe again
when it has evolved to the He subgiant stage. At this stage, the primary
transfers its He-rich envelope stably to the secondary, leading to the
formation of a WD + MS star system. After that, the secondary
continues to evolve and will fill its Roche lobe at the subgiant
stage. At this stage, a CE may be formed due to dynamically
unstable mass transfer. If the CE is ejected, a WD + He star
system could eventually be formed. The subsequent evolution of the
binary is similar to that presented in Figs\,1-2. For this channel,
the primordial binaries are in the range of $M_{\rm 1,i}\sim4.5-7\,\rm
M_{\odot}$, $M_{\rm 2,i}\sim2.5-5\,\rm M_{\odot}$ and $P^{\rm
  i}\sim3-23\,\rm days$.

\subsection{BPS Results}
\subsubsection{Birthrates and delay times of SNe Ia}
\begin{table}
\begin{center}
 \caption{The birthrates and delay times of SNe Ia for three simulation sets based on the violent merger scenario. Notes: $\nu$ and DTDs are the birthrates and the delay time distributions of SNe Ia, respectively.}
 \label{tab5}
   \begin{tabular}{cccccc}
\hline \hline
 Set & $\alpha_{\rm CE}\lambda$ & $\rm\nu (10^{\rm -3}{\rm yr}^{\rm -3})$ & $\rm DTDs ({\rm Gyr})$\\
\hline
$1$ & $0.5$ & $0.01$ & $2.69-8.71$\\
$2$ & $1.0$ & $0.19$ & $>2.14$\\
$3$ & $1.5$ & $0.40$ & $>1.70$\\
\hline \label{1}
\end{tabular}
\end{center}
\end{table}

By performing a series of BPS Monte Carlo simulations, we obtained the
birthrates and delay times of SNe Ia for the violent merger scenario
(see Table\,1). We found that the masses of the primordial primaries
range from $4.0-7.5\,\rm M_{\odot}$, the masses of the primordial
secondaries from $2.5-5.5\,\rm M_{\odot}$, and the primordial
orbital periods are in the range of $3-347\rm\,days$ for producing SNe
Ia based on the WD + He subgiant channel. When these binaries evolve
to become WD + He star systems, the masses of the WDs are in the range of
$0.65-1.07\,\rm M_{\odot}$, the masses of the He stars range from
$0.9-2.1\,\rm M_{\odot}$ and the orbital periods are from
$1.2-10.6\rm\,h$.


Fig.\,9 represents the evolution of the Galactic birthrate of SNe Ia
as a function of time for a constant star-formation rate of $5\,\rm
M_{\odot}yr^{\rm -1}$. This leads to a Galactic SN Ia birthrate for
the violent merger scenario of $\sim0.01-0.4\times10^{\rm -3}\,\rm
yr^{\rm -1}$ based on the WD + He subgiant channel. For comparison,
the observed Galactic SN Ia birthrate is $\sim3-4\times10^{\rm
  -3}\,\rm yr^{\rm -1}$ (e.g. Cappellaro \& Turatto 1997). The violent
merger scenario studied here may thus contribute $\sim0.3\%-10\%$ of
all SNe Ia in our Galaxy. Thus, we suggest that this scenario
is a subchannel for producing SNe Ia. We also note that the SN Ia
birthrate increases with the value of $\alpha_{\rm CE}\lambda$. The
reason is that, for a larger value of $\alpha_{\rm CE}\lambda$, CE
ejection will require less orbital energy, making it easier to form WD
+ He star systems that are located in the correct region for producing
SNe Ia.

\begin{figure}
\begin{center}
\epsfig{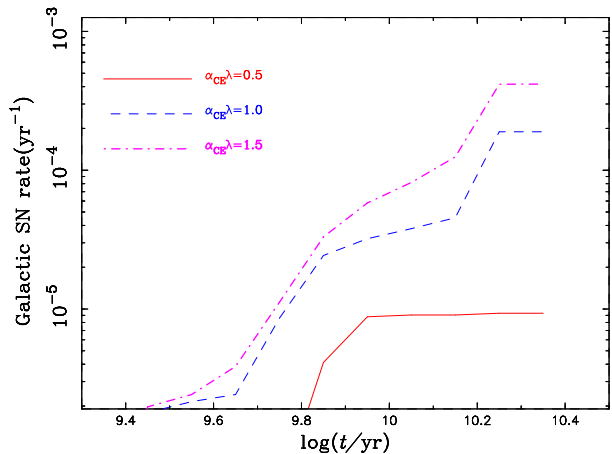}
 \caption{Evolution of Galactic SN Ia birthrates with time for a constant Population I SFR ($Z=0.02$,
${\rm SFR}=5\,\rm M_{\odot}{\rm yr}^{-1}$) based on the WD + He subgiant channel. The solid, dashed and dash-dotted curves show
the results of different CE ejection parameters with $\alpha_{\rm CE}\lambda=0.5$, 1.0 and 1.5, respectively.}
  \end{center}
\end{figure}

In Fig.\,10, we show the delay time distributions for violent mergers
from the WD + He subgiant channel (see thick lines). These
distributions are obtained based on the assumption of a single
starburst producing $10^{10}\,\rm M_{\odot}$ in stars. The delay times
of SNe Ia from the WD + He subgiant channel range from $\sim1.7\,\rm
Gyr$ to the Hubble time. Here, the minimum delay time of ~1.7
Gyrs comes from the left boundary of the grids for producing SNe Ia
shown in Fig. 7. We also note that, if we dropped the mass ratio
constraint, it could be expected that the left boundary of the grids
for producing SNe Ia in Fig. 7 would move to the left, which would
lead to a smaller minimum delay time. We suggest that the violent merger scenario from
this channel may contribute to SNe Ia with long delay times. We also
found that the gravitational wave radiation timescale $t_{\rm GW}$
plays the dominant role in these long delay times. It is worth noting
that there is a real cut-off in $\log(t)$ at large times for the case
with $\alpha_{\rm CE}\lambda=0.5$, while the cut-off is artificial for
the other cases as the system ages have already reached the
Hubble time.

We also note that the DTDs do not obey a $t^{\rm -1}$ relation. There
are two reasons for this, as follows: (i) For
the grids in Fig. 7, the parameter space for producing
violent mergers increases as the orbital period increases, leading
to more binary systems with large orbital periods that can produce SNe
Ia. (ii) The CE ejection process tends to form
binaries with long orbital periods. Since the delay time in this scenario
is sensitive to the orbital period, a binary with long orbital
period will have a long gravitational wave radiation time,
leading to a long delay time.

\begin{figure}
\begin{center}
\epsfig{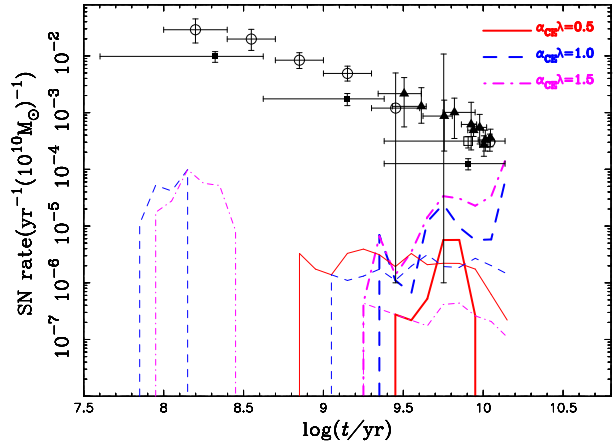}
 \caption{Delay time distributions of SNe Ia based on the violent
   merger scenario. The thick curves show the DTDs of violent mergers
   from the WD + He subgiant channel, while the thin curves represent
   the cases from other channels. The open circles are from Totani et
   al. (2008), the filled triangles, squares are taken from Maoz et
   al. (2010, 2012) and the open square is from Graur \& Maoz (2013).}
\end{center}
\end{figure}

Aside from the WD + He subgiant channel, there are other channels to
form violent mergers (e.g. Ruiter et al. 2011). Those other channels
can also produce violent mergers without going through the particular
mass-accretion process that is so important in this paper (for details
see Meng et al. 2011). Using our BPS code, we also simulated these other
channels for producing SNe Ia via the violent merger scenario. We
found that the Galactic violent merger rate from these channels is
about $0.7-1.8\times10^{\rm -5}\,\rm yr^{\rm -1}$, which is quite
low. The contribution of violent mergers from these channels is less
than 1\% of all SNe Ia (see also Meng et al. 2011). In Fig.\,10, we
also present the DTDs of violent mergers from these other formation channels
(thin curves). The figure shows that the SN Ia delay times
from the violent merger scenario range from $70\,\rm Myr$ to the
Hubble time; violent mergers from the WD + He subgiant channel mainly
contribute to SNe Ia in old populations, while violent mergers from
other channels contribute to SNe Ia in young populations.

\subsubsection{Mass distributions of double WDs}
We calculated the properties of the double WDs from the WD + He
subgiant channel by adopting interpolations in the three-dimensional
grid ($M_{\rm WD}^{\rm i},M_{\rm 2}^{\rm i},\log\,P^{\rm i}$) obtained
in Sect.\,2. We found that the periods of double WDs range from
$1.5\rm\,h$ to $7.5\rm\,h$ at the beginning of their
formation. Fig.\,11 shows the density distribution of violent merger
WDs in the $M_{\rm WD}^{\rm f}-M_{\rm 2}^{\rm f}$ plane. In this
simulation, $4\times10^{\rm 7}$ primordial sample binaries are
included and the CE ejection parameter $\alpha_{\rm CE}\lambda$ is set
to be 1.5. This distribution shows that, in most cases, $M_{\rm
  WD}^{\rm f}$ is larger than $M_{\rm 2}^{\rm f}$, while the fraction
with $M_{\rm 2}^{\rm f}>M_{\rm WD}^{\rm f}$ is only $\sim7\%$. We also
note that the masses of the WDs formed from the primordial primaries
can be as large as $1.3\,\rm M_{\odot}$, which is due to the accretion
of He-rich material from the He donors onto the WDs with lower masses.

\begin{figure}
\begin{center}
\epsfig{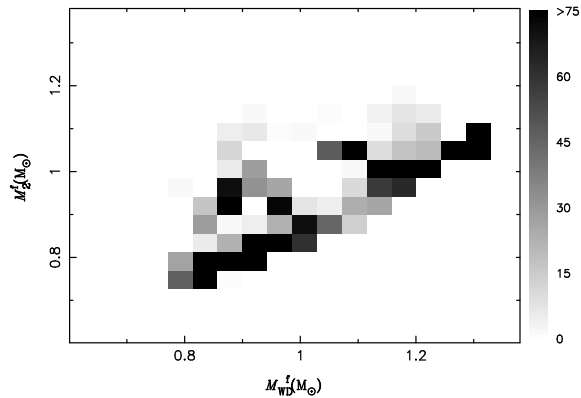}
 \caption{The density distribution of violent mergers from the WD + He
   subgiant channel in the $M_{\rm WD}^{\rm f}-M_{\rm 2}^{\rm f}$
   plane, where $M_{\rm WD}^{\rm f}$ and $M_{\rm 2}^{\rm f}$ are the
   final mass of the primary WD and the mass of the WD produced from
   the He star, respectively. Here, we have set $\alpha_{\rm
     CE}\lambda=1.5$.}
  \end{center}
\end{figure}

In Fig.\,12, we present the initial mass distribution and the final
mass distribution of the primary WDs with $\alpha_{\rm
  CE}\lambda=1.0$. The initial masses of the primary WDs range from
$0.65-1.07\,\rm M_{\odot}$, while the final masses are in the range of
$0.8-1.35\,\rm M_{\odot}$. Note that there are two peaks for these
distributions. The left peak of initial masses is about $0.7\,\rm
M_{\odot}$, while the right peak is around $0.95\,\rm M_{\odot}$; for
the final mass distribution, the left peak is about $0.9\,\rm
M_{\odot}$ and the right peak is near the $1.25\,\rm
M_{\odot}$. Fig.\,13 shows the total mass distribution of the violent
WD mergers for different values of $\alpha_{\rm CE}\lambda$. The
total masses are in the range of $1.4-2.4\,\rm M_{\odot}$. We note
that there are also two peaks in these three distributions. The left
peaks are around $1.7\,\rm M_{\odot}$ and the right ones are about
$2.2\,\rm M_{\odot}$. These two peaks originate from two different
formation channels. The left peaks mainly originate from {\em Channel}
A, while the right ones are mainly from {\em Channel} B (see
Sect.\,3.1); the primordial binaries in {\em Channel} B have slightly
massive primordial primaries and significantly shorter periods, resulting
in the formation of more massive primary WDs.

\begin{figure}
\begin{center}
\epsfig{file=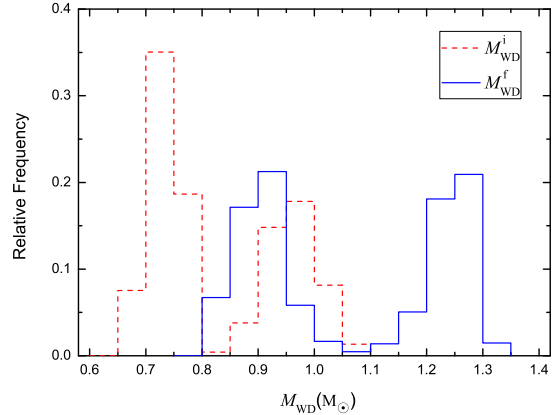,angle=0,width=9.2cm}
 \caption{Distribution of the initial/final masses of the primary WDs
   with $\alpha_{\rm CE}\lambda=1.0$, in which $M_{\rm WD}^{\rm i}$ is
   the initial mass of the primary WDs (red dashed line), and $M_{\rm
     WD}^{\rm f}$ is the final mass of the primary WDs (blue solid
   line).}
  \end{center}
\end{figure}

\begin{figure}
\begin{center}
\epsfig{file=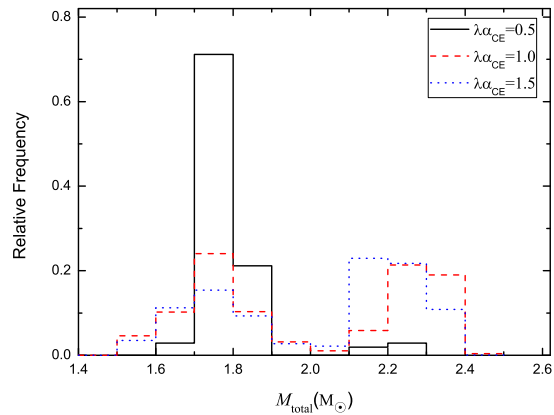,angle=0,width=9.2cm}
 \caption{Distribution of the total mass of double WDs with different
   values of $\alpha_{\rm CE}\lambda$. The black solid, red dashed and
   blue dotted lines represent the cases with $\alpha_{\rm
     CE}\lambda=0.5$, 1.0 and 1.5, respectively.}
  \end{center}
\end{figure}

\section{Discussion}\label{Discussion}

For the violent merger scenario, the mass-accretion process during
which the primary WD accretes material from a He subgiant is crucial
for the production of SNe Ia. By considering an identical value of
$\alpha_{\rm CE}$ as Ruiter et al.\ (2013), i.e. $\alpha_{\rm
  CE}\lambda=0.5$ in this article, we found that such a particular
mass-accretion stage can be expected to occur in about 43.7\% of all
the binary systems which would eventually contribute to SNe Ia; this
is consistent with the fraction given by Ruiter et al.\ (2013). For
other cases with larger $\alpha_{\rm CE}\lambda$, this fraction would
be much higher in our simulations: 91.8\% for the case of $\alpha_{\rm
  CE}\lambda=1.0$ and 85.7\% for the case of $\alpha_{\rm
  CE}\lambda=1.5$. Thus, we suggest that the WD + He subgiant channel
may be the dominant way for producing violent mergers of double WDs
that can form SNe Ia. During the mass-accretion stage, the primary WDs
can increase their mass by about $0.1-0.45\,\rm M_{\odot}$ from the He
donors based on our calculations, which is slightly wider than the
mass range ($\sim0.15-0.35\,\rm M_{\odot}$) presented by Ruiter et
al.\ (2013). This is due to the fact that we employed full stellar
evolution calculations, while Ruiter et al.\ (2013) adopted a simple
analytical fitting formula for estimating the mass-accretion rate. We
compared our prescriptions with Ruiter et al.\ (2013) by calculating
the evolution of a WD + He star system provided in Sect.\,2.3 of
Ruiter et al.\ (2013). This system has a $0.84\,\rm M_{\odot}$ WD and
a $1.25\,\rm M_{\odot}$ He star separated by $1.73\,\rm
R_{\odot}$. Our calculation shows that $M_{\rm WD}^{\rm f}=1.2\,\rm
M_{\odot}$ and $M_{\rm 2}^{\rm f}=0.84\,\rm M_{\odot}$, compared to
$M_{\rm WD}^{\rm f}=1.19\,\rm M_{\odot}$ and $M_{\rm 2}^{\rm
  f}=0.77\,\rm M_{\odot}$ in Ruiter et al.\ (2013). When this double
WDs is formed, their separation is $1.74\,\rm R_{\odot}$ for our
calculation and $1.92\,\rm R_{\odot}$ for Ruiter et al.\ (2013).

In addition, Ruiter et al.\ (2013) claimed that the DTDs from violent
mergers agree rather well with observations and that the violent
merger scenario may be the dominant contributor to SNe Ia in field
galaxies. They argued that violent mergers from the WD + He subgiant
channel may contribute to SNe Ia with delay times $>150\,\rm Myr$, a
conclusion that is quite different from our results. This is
  caused by the different criteria for the critical mass ratio for
  double WDs between Ruiter et al. (2013) and this work. We note that
  Ruiter et al. (2013) used a relatively weak criterion with
  $\eta=1.5$ for the critical mass ratio ($q_{\rm cr}$) of the WD
  binaries to calculate the SN Ia birthrate and delay time
  distribution (see formula\,1 in their paper). From Fig.\,5 of
Ruiter et al.\ (2013), one can see that the difference between
their $q$-cut (with $\eta=1.5$) and no $q$-cut at all is very minor,
especially for the massive WDs of double WDs at the high-mass end,
which means that the double WDs with nearly all mass ratios are
thought to be able to produce SNe Ia in the Ruiter et al.\ (2013)
estimates.


The brightness of SN Ia explosions originating from violent mergers
are mainly determined by the final mass of the massive WDs ($M_{\rm
  WD1}$). The reason is that the less massive WD is totally destroyed
in the merging process and that almost all of the iron-group elements
(including $^{\rm56}\rm Ni$) are produced in the thermonuclear
explosion inside the massive WD. For the violent mergers with low-mass WDs
($M_{\rm WD1}<1.1\,\rm M_{\odot}$), their low density causes
incomplete silicon burning during the explosion, resulting in a lower
yield of $^{\rm56}\rm Ni$ compared with normal SNe Ia,
corresponding to sub-luminous SNe Ia (e.g. Pakmor et al.\ 2010,
2011). However, for the violent mergers with massive WDs ($M_{\rm WD1}>1.1\,\rm
M_{\odot}$), the yields of these massive mergers are still quite
uncertain, and they are potential candidates for explaining
super-Chandrasekhar mass SNe Ia (Moll et al. 2014; Cody et al. 2014).

However, the minimum critical mass ratio of double WDs to produce
prompt detonations and SNe Ia through the violent merger scenario is
still quite uncertain (e.g. Pakmor et al.\ 2011; Sato et
al.\ 2016). In this work, we adopted $q_{\rm cr}\ge0.8$, which is
consistent with Pakmor et al.\ (2011), but somewhat optimistic.
If a larger critical mass ratio (e.g. $q_{\rm cr}\ge0.9$) for
the double WDs is adopted, the contribution of violent mergers to
the Galactic birthrate of SNe Ia would be
$\sim$$0.1$$-$$1.1\times10^{\rm -4}\,\rm yr^{\rm -1}$, accounting
for only 0.2\% to 3\% of all SNe Ia in the Galaxy, and the delay
times of SNe Ia would be larger than 3.5 Gyr. Furthermore, Sato et
al.\ (2015) argued that a SN Ia would be produced during the merging
process if the masses of both WDs are in the range of $0.9\,\rm
M_{\odot} \le M_{\rm WD} \le 1.1\,\rm M_{\odot}$, whereas the
thermonuclear explosion can be expected in the stationary rotating
merger remnant stage if the massive WDs in double WD systems are in
the mass range of $0.7\,\rm M_{\odot} \le M_{\rm WD} \le 0.9\,\rm
M_{\odot}$ and the total masses are larger than the Chandrasekhar mass
(see also Moll et al.\ 2014; Cody et al.\ 2014). Chen et al.\ (2012)
also claimed that the SN Ia birthrate from the DD model may decrease
significantly when considering different constraints for double
WDs. We note that recent studies suggested that a thin He shell on the
surface of CO WDs could potentially produce SNe Ia more easily in
WD mergers (e.g. Dan et al.\ 2012; Pakmor et al.\ 2013).

Furthermore, the merging of CO WD + He WD systems may also contribute
to the birthrates of SNe Ia based on the double-detonation scenario
(see Dan et al.\ 2012; Pakmor et al.\ 2013). In this scenario, mass
transfer is relatively unstable and occurs within the direct impact
regime, in which Kelvin-Helmholz instabilities may trigger an
explosion of the He shell on the surface of the CO WD (see Guillochon
et al.\ 2010). The shock compression in the CO core caused by the He
explosion could potentially lead to the explosion of the whole
WD. Less $^{\rm 56}$Ni would be produced during the thermonuclear
explosions as the accretors always have sub-Chandrasekhar masses; the
SNe Ia resulting from such mergers may have lower luminosities than
normal SN Ia explosions, i.e they may contribute to the class of
sub-luminous SNe Ia (e.g. Sim et al. 2010; Dan et
al. 2012). Considering the possibility of CO WD + He WD systems for
producing sub-luminous SNe Ia, future BPS studies are needed to
explore the properties of SNe Ia through the CO WD + He WD scenario.

\section{Summary}\label{Summary}
In this article, we employed the Eggleton stellar evolution code to
simulate the evolution of WD + He star systems until the onset of the
merging of double WDs, including the effect of an optically thick
wind. With these simulations, we obtained the regions in parameter space
for producing SNe Ia via the violent merger scenario for different
initial WD masses. Using the results from these binary evolution
calculations, we performed a series of BPS Monte Carlo simulations and
obtained the Galactic birthrates and DTDs of SNe Ia. Aside from the WD
+ He subgiant channel, we also obtained the BPS results of violent
mergers from other formation channels. We found that the WD + He
subgiant channel is the dominant contributor of violent mergers and
that the overall SN Ia birthrate from the violent merger scenario
accounts for at most 10 percent of the inferred observational results
in our Galaxy. SNe Ia from the violent merger scenario mainly
contribute to SNe Ia with long delay times based on the WD + He
subgiant channel studied here.  We note that the SN Ia progenitor
survey (SPY) performed by the European Southern Observatory was
designed to search for double WDs (Napiwotzki et al. 2004; Nelemans
et al. 2005; Geier et al. 2007). The double WDs formed from the WD +
He subgiant channel have He-rich atmospheres, which means that those
binaries should be double DB WDs. In order to put further constraints
on the violent merger scenario, large samples of observed double DB
WDs are needed. Additionally, more numerical simulations related to
the violent WD merger scenario are required to constrain the
properties of the resulting SNe Ia.

\section*{Acknowledgments}
We acknowledge useful comments and suggestions from the referee. We acknowledge
Stephen Justham, Xuefei Chen and Xiangcun Meng for their helpful discussions.
This work is supported by the National Basic Research Program of China (973 programme, 2014CB845700),
the National Natural Science Foundation of China (Nos 11322327, 11390374  and 11521303),
the Chinese Academy of Sciences (Nos KJZD-EW-M06-01 and XDB09010202),
the Natural Science Foundation of Yunnan Province (Nos 2013HB097, 2013FB083 and 2013HA005),
and the Youth Innovation Promotion Association CAS.

\label{lastpage}

\begin{thebibliography}{}\label{thebibliography}
\bibitem[Badenes et al. (2007)]{Bad07}           Badenes C., Hughes J. P., Bravo E., Langer N., 2007, ApJ, 662, 472
\bibitem[Barbary et al. (2012)]{Bar12}           Barbary K. et al., 2012, ApJ, 745, 32
\bibitem[Bulla et al. (2016)]{Bul16}             Bulla M. et al., 2016, MNRAS, 455, 1060
\bibitem[Cappellaro \& Turatto (1997)]{Cap97}    Cappellaro E.,  Turatto M., 1997, in Thermonuclear Supernovae, ed. P. Ruiz-Lapuente, R. Cannal, \& J. Isern (Dordrecht: Kluwer), 77
\bibitem[Chakraborti et al. (2015)]{Cha15}       Chakraborti S., Childs F., Soderberg A., 2015, arxiv:1510.08851v1
\bibitem[Chen et al. (2012)]{Che12}              Chen X.., Jeffery C. S., Zhang X., Han Z., 2012, ApJL, 755, 9
\bibitem[Cody et al. (2014)]{Cod14}              Cody R. et al., 2014, ApJ, 788, 75R
\bibitem[Dan et al. (2015)]{Dan15}               Dan M., Guillochon J., Br\"uggen M., Ramirez-Ruiz E., Rosswog S., 2015, MNRAS, 454, 4411
\bibitem[Dan et al. (2012)]{Dan12}               Dan M., Rosswog S., Guillochon J., Ramirez-Ruiz E., 2012, MNRAS, 422, 2417
\bibitem[Dan et al. (2014)]{Dan14}               Dan M., Rosswog S., Br\"ugen M., Podsiadlowski P., 2014, MNRAS, 438, 14
\bibitem[Di Stefano, Voss \& Claeys (2011)]{DV11}  Di Stefano R., Voss R., Claeys J. S. W., 2011, ApJ, 738, L1
\bibitem[Eggleton (1973)]{egg73}                 Eggleton P. P., 1973, MNRAS, 163, 279
\bibitem[Eggleton et al. (2002)]{egg02}          Eggleton P. P., Kiseleva-Eggleton L., 2002, ApJ, 575, 461
\bibitem[Fesen et al. (2015)]{fes15}             Fesen R. A., H\"oflich P. A., Hamilton A. J. S., ApJ, 804, 140
\bibitem[Ganeshalingam, Li \& Filippenko (2011)]{GLF11}  Ganeshalingam M,, Li W., Filippenko A. V., 2011, MNRAS, 416, 2607
\bibitem[Geier et al. (2010)]{Gei10}             Geier S., Heber U., Kupfer T., Napiwotzki R., 2010, A\&A, 515, A37
\bibitem[Geier et al. (2007)]{Gei07}             Geier S., Nesslinger S., Heber U., Przybilla N., Napiwotzki R., Kudritzki R. P., 2007, A\&A, 464, 299
\bibitem[Graham et al. (2015)]{Gra15}            Graham M. L. et al., 2015, MNRAS, 454£¬ 1948
\bibitem[Graur et al. (2011)]{Grau11}            Graur O. et al., 2011, MNRAS, 417, 916
\bibitem[Graur \& Maoz (2013)]{Grau13}           Graur O., Maoz D., 2013, MNRAS, 430, 1746
\bibitem[Guillochon et al. (2010)]{Gui10}        Guillochon J., Dan M., Ramirez-Ruiz E., Rosswog S., 2010,
ApJL, 709, L64
\bibitem[Hachisu, Kato \& Nomoto (1996)]{hac96}  Hachisu I., Kato M., Nomoto K., 1996, ApJ, 470, L97
\bibitem[Hachisu, Kato \& Nomoto (2012)]{hac12}  Hachisu I., Kato M., Nomoto K., 2012, ApJ, 756, L4
\bibitem[Han, Podsiadlowski \& Eggleton (1994)]{HAN94}  Han Z., Podsiadlowski Ph., Eggleton P. P., 1994, MNRAS, 270, 121
\bibitem[Han, Podsiadlowski \& Eggleton (1995)]{HAN95}  Han Z., Podsiadlowski Ph., Eggleton P. P., 1995, MNRAS, 272, 800
\bibitem[Han (1998)]{han98}                      Han Z., 1998, MNRAS, 296, 1019
\bibitem[Han, Tout \& Eggleton (2000)]{han00}    Han Z., Tout C. A., Eggleton P. P., 2000, MNRAS, 319, 215
\bibitem[Han \& Podsiadlowski (2004)]{han04}     Han Z., Podsiadlowski, Ph., 2004, MNRAS, 350, 1301
\bibitem[Hancock et al. (2011)]{Hanc11}          Hancock P. P., Gaensler B. M., Murphy T., 2011, ApJ, 735, L35
\bibitem[Hichen et al. (2007)]{Hic07}            Hicken M. et al., 2007, ApJ, 669, L17
\bibitem[Horesh et al. (2012)]{Hor12}            Horesh A. et al., 2012, ApJ, 746, 21
\bibitem[Howell (2011)]{how11}                   Howell D. A., 2011, Nature Communications, 2, 350
\bibitem[Howell et al. (2006)]{How06}            Howell D. A. et al., 2006, Nature, 443, 308
\bibitem[Hoyel \& Fowler (1960)]{HF60}           Hoyel F., Fowler W. A., 1960, ApJ, 132, 565
\bibitem[Hurley et al. (2002)]{Hur02}            Hurley J. R., Tout C. A., \& Pols O. R., 2002, MNRAS, 329, 897
\bibitem[Iben \& Tutukov (1984)]{IT84}           Iben I., Tutukov A. V., 1984, ApJS, 54, 335
\bibitem[Iben \& Tutukov (1985)]{IT85}           Iben I., Tutukov A. V., 1985, ApJS, 58, 661
\bibitem[Ivanova et al. (2013)]{iva13}           Ivanova N., Justham S., Avendano Nandez J. L., Lombardi J. C., 2013, Science, 339, 433
\bibitem[Justham (2011)]{Jus11}                  Justham S. 2011, ApJL, 730, L34
\bibitem[Kashyap et al. (2015)]{Kps15}           Kashyap R. et al., 2015, ApJ, 800, L7
\bibitem[Kato \& Hachisu (1994)]{kat94}          Kato M., Hachisu I., 1994, ApJ, 437, 802
\bibitem[Kato \& Hachisu (2004)]{kat04}          Kato M., Hachisu I., 2004, ApJ, 613, L129
\bibitem[Kromer et al. (2013)]{Krom13}           Kromer M. et al., 2013, ApJL, 778, L18
\bibitem[Landau \& Lifshitz (1971)]{LL71}        Landau L. D., Lifshitz E. M., 1971, Classical Theory of Fields, Pergamon Press, Oxford
\bibitem[Langer et al. (2000)]{lan00}            Langer N., Deutschmann A., Wellstein S.,  H\"{o}flich P., 2000, A\&A, 362, 1046
\bibitem[Leonard (2007)]{Leo07}                  Leonard D. C., 2007, ApJ, 670, 1275
\bibitem[Li \& van den Heuvel (1997)]{li97}      Li X.-D., van den Heuvel E. P. J., 1997, A\&A, 322, L9
\bibitem[Maoz et al. (2010)]{mao10}              Maoz D., Keren S., Avishay G.-Y., 2010, ApJ, 722, 1879
\bibitem[Maoz et al. (2011)]{mao11}              Maoz D., Mannucci F., Li W., Filippenko A. V., Della Valle M.,  Panagia N., 2011, MNRAS, 412, 1508
\bibitem[Maoz, Mannucci\& Nelemans (2014)]{mao14} Maoz D., Mannucci F., Nelemans G., 2014, ARA\&A, 52, 107
\bibitem[Maoz et al. (2012)]{mao12}              Maoz D., Mannucci F., Timothy D. Brandt, 2012, MNRAS, 426, 3282
\bibitem[Maxted et al. (2000)]{Max00}            Maxted P. F. L., Marsh T. R., North R. C., 2000, MNRAS, 317, L41
\bibitem[Meng et al. (2011)]{men11}              Meng X., Chen W., Yang W., Li Z., 2011, A\&A, 525, A129
\bibitem[Meng et al. (2009)]{men09}              Meng X., Chen X., Han Z., 2009, MNRAS, 395, 2103
\bibitem[Mennekens et al. (2010)]{menn10}        Mennekens N., Vanbeveren D., De Greve J. P., De Donder E., 2010, A\&A, 515, A89
\bibitem[Miller \& Scalo (1979)]{mil79}          Miller G. E., Scalo J. M., 1979, ApJS, 41, 513
\bibitem[Moll et al. (2014)]{Mol14}              Moll R., Raskin C., Kasen, D., Woosley S. E., 2014, ApJ, 785, 105M
\bibitem[Napiwotzki et al. (2004)]{Nap04}        Napiwotzki R., Yungelson L., Nelemans G. et al., 2004. ASPC 318, 402
\bibitem[Napiwotzki et al. (2007)]{Nap07}        Napiwotzki R. et al., 2007, ASPC, 372, 387
\bibitem[Nelemans et al. (2005)]{Nel05}          Nelemans G., Napiwotzki R., Karl C. et al., 2005. A\&A 440, 1087
\bibitem[Nelemans et al. (2001)]{Nel01}          Nelemans G., Yungelson L. R., Portegies Zwart S. F., Verbunt F., 2001, A\&A, 365, 491
\bibitem[Nomoto (1982)]{nom82}                   Nomoto K., 1982, ApJ, 253, 798
\bibitem[Nomoto \& Iben (1985)]{nom85}           Nomoto K., Iben I., 1985, ApJ, 297, 531
\bibitem[Pakmor et al. (2010)]{Pak10}            Pakmor R., Kromer M., R\"{o}pke F. K., Sim S. A., Ruiter A. J., Hillebrandt W., 2010, Nature, 463, 61
\bibitem[Pakmor et al. (2011)]{Pak11}            Pakmor R., Hachinger S., R\"{o}pke F. K., Hillebrandt W., 2011, A\&A, 528, A117
\bibitem[Pakmor et al. (2012)]{Pak12}            Pakmor R. et al., 2012, ApJ, 747, L10
\bibitem[Pakmor et al. (2013)]{Pak13}            Pakmor R., Kromer M., Taubenberger S., Springel V., 2013, ApJL, 770, L8
\bibitem[Perlmutter et al. (1999)]{per99}        Perlmutter S. et al., 1999, ApJ, 517, 565
\bibitem[Podsiadlowski et al. (2008)]{Pod08}     Podsiadlowski Ph., Mazzali P., Lesaffre P., Han Z., F\"orster F., 2008, New Astro. Rev., 52, 381
\bibitem[Pols et al. (1998)]{pol98}              Pols O. R., Schr\"{o}der K. P., Hurly J. R., Tout C. A., Eggleton P. P., 1998, MNRAS, 298, 525
\bibitem[Pols et al. (1995)]{pol95}              Pols O. R., Tout C. A., Eggleton P. P., Han Z., 1995, MNRAS, 274, 964
\bibitem[Riess et al. (1998)]{rie98}             Riess A. et al., 1998, AJ, 116, 1009
\bibitem[Rodr\'iguez-Gil et al. (2010)]{Rod10}   Rodr\'iguez-Gil P. et al., 2010, MNRAS, 407, L21
\bibitem[R\"{o}pke et al. (2012)]{rop12}         R\"{o}pke F. K. et al., 2012, ApJL, 750, L19
\bibitem[Ruiter et al. (2011)]{Rui11}            Ruiter A. J., Belczynski K., Sim S. A., Hillebrandt W., Fryer C. L., Fink M., Kromer M., 2011, MNRAS, 417, 408
\bibitem[Ruiter et al. (2009)]{Rui09}            Ruiter A. J., Belczynski K., Fryer C., 2009, ApJ, 699, 2026
\bibitem[Ruiter et al. (2013)]{Rui13}            Ruiter A. J. et al., 2013, MNRAS, 429, 1425
\bibitem[Saio \& Nomoto (1985)]{Sai85}           Saio H., Nomoto K., 1985, A\&A, 150, 21
\bibitem[Saio \& Nomoto (1998)]{Sai98}           Saio H., Nomoto K., 1998, ApJ, 500, 388
\bibitem[Sand et al. (2012)]{San12}              Sand D. J. et al., 2012, ApJ, 746, 163
\bibitem[Sato et al. (2015)]{Sat15}              Sato Y. et al., 2015, ApJ, 807, 105S
\bibitem[Sato et al. (2016)]{Sat16}              Sato Y. et al., 2016, ApJ, arxiv:1603.01088v1
\bibitem[Scalzo et al. (2010)]{Sca10}            Scalzo R. A. et al., 2010, ApJ, 713, 1073
\bibitem[Seitenzahl et al. (2015)]{Seit15}       Seitenzahl I. R. et al. 2015, MNRAS, 447, 1484l
\bibitem[Sim et al. (2010)]{Sim10}               Sim S. A., R\"{o}pke F. K., Hillebrandt W., Kromer M., Pakmor R., Fink M., Ruiter A. J., Seitenzahl I. R., 2010, ApJL, 714, L52
\bibitem[Taubenberger et al. (2013)]{Tau13}      Taubenberger S. et al., 2013, 775, L43
\bibitem[Tanikawa et al. (2015)]{Tan15}          Tanikawa A. et al., 2015, ApJ, 807, 40
\bibitem[Timmes et al. (1994)]{Tim94}            Timmes F. X., Woosley S. E., Taam Ronald E., 1994, ApJ, 420, 348T
\bibitem[Toonen et al. (2012)]{Too12}            Toonen S., Nelemans G., Portegies Zwart S., 2012, A\&A, 546, A70
\bibitem[Totani et al. (2008)]{tot08}            Totani T., Morokuma T., Oda T., Doi M., Yasuda, N., 2008, PASJ, 60, 1327
\bibitem[Tovmassian et al. (2010)]{Tov10}        Tovmassian G. et al., 2010, ApJ, 714, 178
\bibitem[Umeda et al. (1999)]{Ume99}             Umeda H., Nomoto K., Yamaoka H., Wanajo S., 1999, ApJ, 513, 861
\bibitem[Wang \& Han (2010)]{wh10}               Wang B., Han Z., 2010, A\&A, 515, A88
\bibitem[Wang, Li \& Han (2010)]{WLH10}          Wang B., Li X.-D., \& Han Z., 2010, MNRAS, 401, 2729
\bibitem[Wang \& Han (2012)]{wh12}               Wang B., Han Z., 2012, New Astron. Rev., 56, 122
\bibitem[Wang et al. (2014)]{Wan14}              Wang B., Justham S., Liu Z.-W., Zhang J.-J., Liu D.-D., Han Z., 2014, MNRAS, 445, 2340
\bibitem[Wang et al. (2009)]{Wan09}              Wang B., Meng X., Chen X., Han Z., 2009, MNRAS, 395, 847
\bibitem[Webbink (1984)]{web84}                  Webbink R. F., 1984, ApJ, 277, 355
\bibitem[Whelan \& Iben (1973)]{whe73}           Whelan J., Iben I., 1973, ApJ, 186, 1007
\bibitem[Willems \& Kolb (2004)]{Will04}         Willems B., Kolb U., 2004, A\&A, 419, 1057
\bibitem[Yoon et al. (2007)]{Yoon07}             Yoon S.-C., Podsiadlowski Ph., Rosswog S., 2007, MNRAS, 390,933
\bibitem[Yungelson \& Livio(1998)]{YL98}         Yungelson L., Livio M., 1998, ApJ, 497, 168
\end{thebibliography}
\end{document}